# Cosmologia e Representação[1]

*Marcelo Byrro Ribeiro*[2]

Instituto de Física

Universidade Federal do Rio de Janeiro – UFRJ

## 1. Introdução

*De onde viemos? Para onde vamos? Qual a origem de tudo que nos cerca? Houve um início para esse todo, para esse Universo, que nos cerca? Se houve um início, como e quando ocorreu?* Essas são perguntas têm sido feitas desde tempos imemoriais, em qualquer cultura e em qualquer época. O conjunto de respostas a essas perguntas, que necessariamente incluirão implicitamente algum tipo de definição acerca de noções como "todo", "origem" e "Universo" constituem uma **cosmologia**. Se essas questões forem respondidas em um contexto de mitos de criação, então teremos um conjunto de respostas que constituem uma cosmologia mitológica. Já se essas respostas estiverem dentro de um contexto baseado na crença aos Deuses, ou em apenas um só Deus, teremos uma cosmologia religiosa. Vivemos em uma sociedade onde a ciência e a tecnologia têm um papel fundamental, formando um dos pilares essenciais da civilização moderna. Assim, é natural que as perguntas acima também sejam formuladas dentro de um contexto científico, o que significa que a sociedade moderna também foi capaz de produzir uma cosmologia, porém baseada na ciência e tecnologia. Tal *cosmologia científica* foi gerada após um longo processo de desenvolvimento e amadurecimento da ciência moderna, ocorrido durante séculos, até que esta alcançou

---

[1] Palestra apresentada na mesa redonda *"O Debate entre Ciência e Religião"* ocorrido no XII Simpósio Nacional da Associação Brasileira de História das Religiões (ABHR) em 2 de junho de 2011 na Universidade Federal de Juiz de Fora.

[2] E-mail: mbr at if.ufrj.br; Internet page: http://www.if.ufrj.br/~mbr/

um grau de desenvolvimento onde as ciências físicas puderam formular as questões acima dentro de seus próprios critérios científicos e responder, pelo menos parcialmente, essas perguntas. Esse conjunto de respostas e a atividade de responder às perguntas cujas respostas ainda não são inteiramente satisfatórias, dentro dos critérios de qualidade da pesquisa científica moderna, formam atualmente uma área de pesquisa de intensa atividade conhecida como *cosmologia científica moderna*.

O objetivo desse trabalho é apresentar uma breve descrição da cosmologia científica moderna, seus resultados principais e suas bases conceituais mais gerais. Procurarei fazer isso dentro de um arcabouço epistemológico que permita um diálogo com outras formas de pensamento, em particular o religioso, mas sem se restringir a este. Resumidamente, essa visão epistemológica propõe que *as teorias científicas são representações, imagens, da natureza*. Dentro desse ponto de vista a ciência não pode conhecer as essências ou, o que é o mesmo, porque o mundo é tal como ele é realmente. Assim, as respostas propostas pela ciência moderna são parciais, simplificadas e substituíveis. Uma outra forma de ver isso é afirmar que *a verdade é provisória*. E, como consequência natural dessa visão, o mesmo conjunto de fenômenos, ou perguntas científicas, podem ter múltiplas respostas, ou representações, conclusão esta comumente denominada de *pluralismo teórico*. Portanto, esse pluralismo de concepções da natureza abre espaço para o diálogo entre diferentes formas de pensamento.

Não há aqui qualquer pretensão de se chegar a respostas finais às perguntas cosmológicas apresentadas no início desse texto. Alias, como dito acima, e veremos em detalhes abaixo, o arcabouço epistemológico proposto exclui essa possibilidade, ou seja, não permite que possamos jamais chegar a respostas finais, ou últimas, para qualquer problema, seja ele científico ou não.  No entanto, o espaço aberto para o diálogo permite que ambos os lados possam sair enriquecidos desde que as ideias geradas nesse diálogo mostrem-se úteis dentro dos domínios de validade de cada forma de pensamento uma vez que sejam averiguadas e aprovadas pelos critérios de qualidade próprios de cada domínio. Portanto, não há aqui qualquer preocupação de se criticar a religião em nome da ciência ou vice-versa. Os tempos de beligerância entre ciência e religião devem ser deixados para trás.

## 2. Cosmologia Científica Moderna

Durante cerca de quatro séculos a ciência moderna rotulou como não-científicas as questões que indagavam pela origem de tudo, pela história e pelo desenvolvimento dos objetos e estruturas celestes. Afinal era preciso se distinguir radicalmente da filosofia e da teologia e conquistar a sua autonomia e independência com relação a essas duas formas mais antigas de conhecimento. Mas, principalmente, faltavam à ciência moderna os conceitos e equações físicas apropriadas que permitissem abordar e desenvolver as ideias de "todo" e "origem" segundo os seus próprios critérios. Por não possuir as ferramentas necessárias para lidar com essas questões, ou seja, leis e teorias científicas acopladas aos essenciais dados astronômicos, a ciência moderna não estava suficientemente madura para se aventurar na busca de respostas para o conjunto de questões cujas respostas definem uma cosmologia: de onde viemos? Para onde vamos? Qual é a origem de tudo aquilo que nos cerca? Qual é a nossa origem? Qual será o nosso fim?

A maturidade da ciência é uma tarefa que envolve muitas gerações de pesquisadores, ou seja, nunca é alcançada facilmente, e inclui teorias sólidas baseadas em experimentos reprodutíveis e observações confiáveis. No domínio da cosmologia esses resultados foram obtidos pela física somente no século XX após Albert Einstein (1879-1955) ter publicado em 1916 a formulação final para a sua *Teoria da Relatividade Geral* (TRG), a qual permitiu que a ciência moderna pudesse estabelecer uma cosmologia dentro de seus próprios termos. Por meio dela pode-se definir o Universo físico como a *totalidade* daquilo que podemos observar com nossos telescópios e satélites artificiais, do que não podemos observar agora por esses meios, mas poderemos no futuro, e daquilo que jamais poderemos observar. A cosmologia científica moderna define conceitos próprios para início, criação, evolução e horizonte, estabelece que o Universo está submetido à força gravitacional cujas equações descritivas e resultados de suas previsões teóricas, obtidas por meio da TRG, são passíveis de verificação e validação empírica por meio da astronomia. Como consequência dessa pesquisa, um dos mais significativos resultados da cosmologia do século XX diz respeito ao fato de que *o Universo evolui*, ou seja, que ele tem uma *história*.

A ideia de evolução tornou-se científica em um domínio estranho à cosmologia científica moderna. Ela já estava presente em outras áreas do saber humano, como a biologia, a história e a antropologia, em uma época em que a cosmologia científica ainda não integrava o domínio das disciplinas científicas estritas, ou seja, nos tempos de Charles Darwin (1809-1882) na segunda metade do século XIX. Mas, para penetrar na cosmologia e receber sua *interpretação* em termos cosmológicos, a ideia de evolução teve que superar resistências iniciais, algumas destas formuladas pelo próprio Albert Einstein, ou seja, de um dos principais responsáveis pela transformação da cosmologia em ciência. Contrariando as equações da sua então recentemente proposta TRG e seguindo as suas ideias científicas preconcebidas até aquele momento, Einstein propôs em 1917 um universo estático. Anos mais tarde ele abandonou a solução do universo estático afirmando aquele ter sido o seu maior erro científico.

Conforme discutido acima, como as questões que definem a cosmologia científica moderna não lhe pertencem inteiramente ou unicamente, é natural que habitantes de outras áreas de conhecimento desejem respondê-las. Refiro-me aqui aos filósofos e teólogos. Se a ciência moderna sente-se suficientemente segura e madura para, dentro de seus próprios critérios, responder às perguntas cosmológicas, por que haveria ela, então, sentir qualquer receio de dialogar com a religião e a teologia? A maturidade e a confiança da ciência moderna, não raramente defendidos de forma arrogante, devem se traduzir em vontade e coragem de conversar com outrem. O diálogo certamente fica mais rico e interessante quando se reconhece a importância de ouvir aqueles que habitam outros domínios científico epistemológicos. Afinal, em que ocasião afirmamos mais fundamentalmente nossa humanidade do que quando nos perguntamos pela nossa origem e pelo nosso destino?

## 3. A Física do Universo

A cosmologia científica nasceu em 1917 pois nesse ano Einstein publicou um artigo no qual as suas recém-propostas equações do campo gravitacional obtidas na TRG foram usadas para estudar a física do Universo. Neste artigo o físico alemão assumiu que *o Universo pode ser tratado como um objeto único*, uma entidade física única, e que o estudo do *Universo como um todo* é possível de ser feito por meio das leis da física. Basicamente,

Einstein estabeleceu o problema cosmológico moderno e apresentou as duas perguntas essenciais que até hoje definem o objetivo da pesquisa em cosmologia científica, que é *obter a estrutura geométrica e a distribuição de matéria do universo*. As respostas dadas por Einstein a essas duas perguntas em seu artigo de 1917 foram insatisfatórias e hoje são consideradas inteiramente ultrapassadas e obsoletas. No entanto, embora as respostas atuais sejam bastante diferentes, as perguntas são ainda essencialmente as mesmas.

Além de formular as perguntas básicas, Einstein fez outra contribuição fundamental à cosmologia, pois somente foi possível iniciar e continuar este estudo porque foram formuladas, implícita ou explicitamente, várias hipóteses fundamentais as quais permitem, e até certo ponto definem, o escopo da cosmologia científica.

## 3.1 As Hipóteses Cosmológicas

A *primeira* hipótese assume que as leis físicas verificáveis dentro dos limites do Sistema Solar são igualmente válidas em regiões muitos mais distantes nas quais esta validade não foi verificada ou onde não podemos testá-las com os meios tecnológicos atuais. Assim, a cosmologia científica moderna estuda todos os objetos e grupos de objetos físicos, incluindo os mais remotos, nos quais as nossas leis físicas têm significado e podem ser aplicadas de maneira consistente e bem-sucedida. A justificativa desta hipótese deve ser observacional, ou seja, a astronomia e astrofísica devem nos fornecer dados que possam justificar esta premissa. Dados obtidos por meio da cuidadosa, paciente e sistemática observação dos céus feita por gerações de pesquisadores usando telescópios e, mais recentemente, satélites artificiais, nos permitem validar esta hipótese em termos gerais.

A *segunda* premissa básica é a de que os objetos cosmológicos interagem gravitacionalmente. Isto é consequência da constatação de que das quatro forças fundamentais conhecidas pela física, a saber, as forças gravitacional, eletromagnética e nucleares forte e fraca, somente a gravidade e a força eletromagnética têm longo alcance, mas considerando a evidência empírica de que os objetos celestes são eletricamente neutros, a gravidade é, portanto, a interação física mais importante entre os objetos cosmológicos. A teoria gravitacional cujos conceitos e equações permitem lidar com conceitos com o "todo" ou "totalidade" é a TRG. Portanto, não é coincidência o fato de que a cosmologia efetivamente nasce como disciplina científica somente *após* o aparecimento da TRG, a qual permitiu Einstein pensar o Universo em termos físicos e usar grandezas físicas típicas como energia e densidade para descrever o Universo.

A *terceira* hipótese cosmológica é na verdade um conjunto de *pressuposições físico-matemáticas* que na prática têm o objetivo de simplificar matematicamente o problema a ser resolvido. Elas, portanto, têm um caráter subjetivo e expressam opiniões filosóficas subjacentes, sendo fonte freqüente de controvérsias devido a ausência de validação observacional direta, tendo, na melhor das hipóteses, indicação indireta. No entanto, elas são essenciais pois sem elas não seria possível obter soluções das equações do campo gravitacional fornecidas pela teoria gravitacional subjacente. As equações da TRG são bastante complexas e de difícil solução, necessitando, portanto, de serem simplificadas matematicamente.

A primeira destas pressuposições matemáticas consiste na adoção do *princípio Copernicano* o qual afirma que não existem observadores privilegiados no Universo. Esse princípio é transcrito matematicamente na forma do assim chamado *princípio cosmológico*, o qual, na prática, consiste em uma justificativa da *escolha*, entre várias possíveis, da geometria do universo como sendo a de Friedmann-Lemaître-Robertson-Walker[3] (FLRW).

A segunda destas pressuposições assume que o Universo pode ser modelado como um fluido sendo as galáxias seus elementos constituidores, isto é, elas são as "moléculas" desse fluido. Pode-se, portanto, descrever esse *fluido cosmológico* por meio de grandezas físicas típicas como densidade, pressão e temperatura.

Há hoje um grande conjunto de evidências astronômicas associadas ao modelo FLRW (geometria FLRW + fluido cosmológico) a ponto de este ser chamado genericamente de

---

[3] Alexander A. Friedmann (1888-1925) foi o matemático russo que primeiro propôs, em 1922 e 1924, soluções das equações da TRG nas quais o Universo estava em expansão. O mesmo tipo de solução com expansão foi independentemente proposta pelo físico, matemático e padre belga Georges E. Lemaître (1894-1966) em 1927 e 1929. Lemaître também analisou as implicações físicas e chegou à conclusão de que poderia ter existido um *átomo primordial* a partir do qual o Universo teria se expandido. Em 1935 o matemático estadunidense Howard P. Robertson (1903-1961) e o matemático inglês Arthur G. Walker (1909-2001) demonstraram independentemente que as soluções de Friedmann e Lemaître eram as mais gerais compatíveis com uma determinada classe de soluções matemáticas.

*modelo cosmológico padrão*. Apesar de existirem modelos com outras geometrias, essas usam o modelo FLRW como referência.

## 3.2 O Modelo Cosmológico Padrão

A primeira característica importante do modelo padrão é que o Universo encontra-se em expansão, ou seja, é *dinâmico*, e seu conteúdo material muda com o tempo. Em outras palavras, o modelo evolui. A dinâmica do universo pode ser determinada por três submodelos:

1. *submodelo aberto*: o universo encontra-se em expansão eterna;
2. *submodelo plano*: o universo também se expande eternamente, mas somente com a energia mínima necessária para tal, enquanto que no caso acima a energia para manter a expansão eterna é acima da mínima;
3. *submodelo fechado*: a expansão do universo vai decrescendo até que ele atinge um máximo. A partir daí o universo inicia uma contração.

O que distingue esses três casos é a quantidade de matéria no universo. Esses submodelos definem o conceito de *massa crítica*, acima da qual o universo vai finalmente paralisar sua expansão e iniciar um processo de contração, e abaixo da qual o universo se expande eternamente. O caso intermediário, denominado universo plano, define a massa crítica: se o universo tiver massa total inferior ao do modelo plano, então é aberto. Se for superior, é fechado. Uma vigorosa área de pesquisa em cosmologia tem sido a tentativa de determinar qual dos três tipos mais se aproxima ao universo observado. Se assumirmos, por argumentos teóricos, que o universo é plano e, ao mesmo tempo, observamos matéria compatível somente com o modelo aberto, uma maneira de resolver esta contradição é supormos que existe matéria invisível, que não emite luz. Em outras palavras, esse cenário implica que há grande quantidade de *matéria escura*.

Observações astronômicas realizadas na última década do século XX sugerem que o universo pode estar se *acelerando*, ou seja, a matéria estaria se afastando mais rapidamente entre si do que o previsto no submodelo aberto. Para que isso ocorra dentro do cenário do modelo FLRW é preciso avançar a hipótese da existência de uma força de expansão extra, proveniente de algum tipo de matéria não detectável diretamente e de natureza desconhecida. Essa matéria interagiria gravitacionalmente, gerando uma energia de aceleração proveniente desse material ainda não detectado. Tal energia foi batizada de *energia escura*. Uma vigorosa pesquisa teórica e observacional, envolvendo vários telescópios espalhados pelo mundo e o lançamento de satélites artificiais dedicados, se iniciou na primeira década do século XXI com o objetivo de procurar rastrear a natureza e composição dessa possível energia escura.

Uma conclusão imediata que emerge da constatação da existência da expansão é que o universo deve ter estado "concentrado" no passado, tendo, por motivos não esclarecidos, se iniciado um processo de expansão em determinada época. Este é o átomo primordial sugerido por Lemaître. Deve-se enfatizar que o átomo primordial não necessariamente implica em um ponto ou uma "origem", pois o universo poderia ter estado em estado estático, sem contração ou expansão, até que algum evento tenha dado início a uma expansão. No entanto, cálculos posteriores mostraram que os três sub-modelos FLRW, apresentados acima, implicam que esta concentração de matéria teria ocorrido em um tempo finito do passado, onde toda a matéria do universo concentrada e com densidade e temperaturas infinitas. Desta constatação matemática nasce o conceito da "grande explosão", ou *big bang*, que é interpretado por muitos como sendo a *criação do Universo*. No caso dos modelos fechados, se em algum momento o universo parar a expansão e iniciar um processo de contração, então quando o universo estiver todo contraído de volta a um ponto temos o que se chama de "grande implosão", ou *big crunch*.

É importante observar que o *big bang* e o *big crunch* são noções *matemáticas*. O modelo teórico prevê estes estados como casos limites, isto é, se o processo de evolução do Universo continuasse, e isto sob o implacável ponto de vista da lógica. Porém, a teoria não é capaz de dizer nada sobre o que eles significam ou como e porque eles podem ter aparecido. A própria física não pode ser feita *no big bang*, pois não se conhece nenhum sistema físico que tenha densidade e pressão infinitos. Associar ao big bang uma idéia de "criação" é, na verdade, uma interpretação com raízes, filosóficas. Teorias baseadas na física quântica aplicadas à cosmologia, a chamada *cosmologia quântica*, tentam superar essas limitações, mas até o momento tal abordagem não produziu resultados satisfatórios.

## 3.3 Evidências Observacionais do Modelo Cosmológico Padrão

Conforme discutido acima, o modelo cosmológico padrão estabelece que *(1) o Universo está em expansão*. Tal foi comprovado empiricamente por Lemaître em 1927 e pelo astrônomo norte-americano Edwin P. Hubble (1889-1953) em 1929. A descoberta da expansão do universo liquidou de vez com a credibilidade do primeiro modelo cosmológico estático proposto por Einstein em 1917, que hoje tem valor apenas histórico. A comprovação da expansão sugere *(2) a evolução do Universo*, resultado consistente com as observações de galáxias distantes em diferentes estágios de desenvolvimento.

Os dados também nos mostram que *(3) o Universo é isotrópico*. Isotropia significa igualdade em todas as direções, ou seja, não existe diferença observacional ou teórica entre duas direções distintas. A contagem de galáxias em diferentes regiões do céu nos mostra empiricamente que não existe nenhuma direção privilegiada e o fato de existir uma radiação de fundo na frequência das micro-ondas, associada aos estágios iniciais do universo, a qual é detectada isotropicamente, fornece uma segunda evidência para a isotropia do universo.

Dois grupos de evidências astronômicas também indicam que *(4) o Universo passou por uma fase densa e quente*. A primeira evidência é a própria radiação de micro-ondas, cuja melhor explicação é sua origem cosmológica e indica que a temperatura esteve mais alta no passado e, portanto, o universo foi mais quente. De acordo com o modelo FLRW, temperaturas mais altas estão vinculadas a maiores densidades, devido ao efeito reverso da expansão, isto é, no passado, o universo estava mais concentrado. A segunda fonte de evidência é a composição química da Terra, dos meteoritos, do sistema solar, das estrelas distantes e das galáxias. Esta composição é bem explicada pela existência de fusões nucleares em uma bola de fogo primordial, as quais produzem genericamente as abundâncias químicas observadas. Assim, a composição química observada seria resultado do processamento nuclear da matéria em uma época densa e quente. Por meio desse resultado pode-se estabelecer a *(5) história térmica do universo* a qual descreve as diversas fases de expansão e resfriamento.

Há, finalmente, alguma evidência de que *(6) o Universo é homogêneo*. Métodos indiretos, provenientes da análise dos dados obtidos em telescópios e satélites artificiais, sugerem esta evidência.

# 4. Conceitos Fundamentais da Cosmologia Científica

Quando em 1917 Einstein assumiu que o Universo poderia ser tratado como um objeto único e empregou suas recém propostas equações do campo gravitacional obtidas na TRG para estudar a física do Universo, surgiu então uma nova disciplina científica, a *cosmologia*. Unicidade e totalidade passam a integrar o vocabulário da física, pois foi a primeira vez que a ideia de *totalidade*, presente até aquele momento apenas nas cosmologias mitológicas e filosóficas, ganhou uma consistente e coerente interpretação físico-matemática. O Universo transforma-se então em um objeto físico, passível de ser descrito e estudado por meio de grandezas físicas típicas como energia, pressão, densidade, temperatura, entre outras, e conceitos e métodos matemáticos e geométricos.

Houve tentativas similares anteriores a Einstein, mas essas esbarraram em dificuldades, na época insuperáveis, devido a notória dificuldade da teoria gravitacional proposta por Isaac Newton (1642-1727), cerca de 250 anos antes de Einstein, em lidar com sistemas infinitos. O universo físico é, por definição, infinito, e isso requer uma teoria física capaz de lidar matematicamente com essa característica. A TRG provê essa teoria, permitindo finalmente que a interpretação física de ideias como "Universo" e "totalidade" sejam colocadas em bases físicas seguras.

## 4.1 Universo e Totalidade

A cosmologia científica moderna entende o conceito de totalidade como sendo o Universo, isto é, como o conjunto de todos os objetos que nos cercam e se influenciam mutuamente. Esse conjunto é *como* uma entidade física única, a qual pode ser descrita através de conceitos e métodos da geometria diferencial, cujos fundamentos foram estabelecidos por Georg F. B. Riemann (1826-1866), e variáveis físicas típicas, tais como densidade, energia e pressão. Essas últimas têm seu comportamento regido pelas soluções das equações do campo gravitacional oriundas da TRG.

A conceituação desse universo físico se desdobra em dois pilares conceituais fundamentais, os quais, ao mesmo tempo em que o definem, o descrevem e também o sustentam. O primeiro é o *empírico/observacional*, isto é, aquilo que relaciona o universo físico com a natureza. O segundo é o *representativo*, onde se encontram a matemática e as teorias físicas subjacentes. Mais especificamente, podemos dizer que o universo físico da cosmologia moderna apoia-se sobre as seguintes bases:

- Aquilo que podemos observar astronomicamente nos céus e relacionar experimentalmente com a física conhecida e descrita nos laboratórios terrestres. Por exemplo, quando captamos e estudamos por meio de instrumentos especiais a luz das galáxias distantes podemos concluir se existe ou não algum elemento conhecido na Terra, como o cálcio ou o lítio, apenas comparando as propriedades da luz emitidas por estes mesmos elementos em laboratórios terrestres;
- Aquilo que podemos concluir quando usamos as teorias físicas conhecidas, em particular a TRG, e a sua subjacente expressão em linguagem e conceitos matemáticos. Por exemplo, ao usarmos a *geometria Riemanniana*, a qual define e descreve matematicamente o conceito de curvatura, junto à hipótese relativística de que o espaço e o tempo são indissociáveis, obtemos uma descrição geométrica do espaço-tempo como tendo quatro dimensões (uma temporal e três espaciais) onde podemos falar de uma curvatura neste mesmo espaço-tempo. É esta curvatura quadri-dimensional que define gravidade na TRG.

Estes dois aspectos são fundamentais, intrínsecos e indissociáveis na cosmologia moderna. A parte observacional diz que o Universo natural mais amplo possível deve ser incluído no universo físico, e por mais amplo possível se deve entender o mais *distante* possível. Deve-se notar aqui que o termo "distância" deve ser entendido no seu sentido relativístico, ou seja, distância espacial e "distância" temporal. Isso significa que tanto os objetos situados longe de nós quanto os do nosso passado remoto fazem parte da totalidade cosmológica.

Do que foi dito acima, conclui-se que não é possível haver uma distinção clara entre os conceitos físicos e os correspondentes *modelos de universo*. Os conceitos físicos, são necessários para se conceber e entender o universo físico. As teorias físicas tornam-se então irremediavelmente mescladas com as suas correspondentes aplicações, o que implica que no caso da cosmologia aquilo que entendemos por *universo observável* não pode existir independente de uma construção teórica. Antes do aparecimento da TRG não havia cosmologia, isto é, não havia uma disciplina científica que discutia *a física do Universo*. Embora o Universo exista independentemente de nós, só se tornou possível sondá-lo e estudá-lo por meio de uma interpretação teórica. Assim, em cosmologia, como em toda e qualquer ciência, é a teoria que determina o que é observacional, ou mesmo o que é observável. Ao mesmo tempo, a *cosmologia observacional* sugere que partes da teoria necessitam de atualização e revisão. Portanto, em cosmologia, teoria e observação interagem em mão dupla, uma modificando, concebendo e limitando a outra. O conceito de horizonte mostra de forma enfática como em cosmologia se faz presente esta vinculação de definição e limitação entre teoria e observação.

## 4.2 Horizontes e Universo Observável

Um dos postulados básicos da TRG afirma que a velocidade da luz é constante. A velocidade da luz, embora muito alta, não é infinita. Assim, a transmissão de um sinal deve ocorrer em um tempo finito, para todos os observadores, pois todos medem o mesmo valor para esta velocidade. Há aqui uma consequência importante. Informação que chegue a nós de uma estrela ou galáxia distante (por exemplo, uma explosão) demorará para nos atingir, e enquanto ela não chegar nós não teremos nenhuma *informação*, a qual é "carregada" pela luz, acerca dessa galáxia ou estrela. Em outras palavras, algo poderá ter ocorrido do outro lado do Universo, mas que demorará algum tempo, talvez muito tempo, até que percorra o caminho até nós e possamos então tomar conhecimento da existência deste fenômeno (por exemplo, a explosão). No jargão técnico da relatividade, diz-se que no momento desse evento (explosão) esta estrela ou galáxia estava fora de nosso *horizonte*.

Os horizontes são uma consequência da finitude da velocidade da luz, e implicam que nós não temos conhecimento imediato de tudo o que ocorre no Universo. Só saberemos da existência de determinados fenômenos em nosso futuro. Esta é uma limitação *teórica* intrínseca. Se, no entanto, a cosmologia concebe o Universo como um todo, a totalidade, e ao mesmo tempo concebe a existência de horizontes, é inescapável a conclusão de que, ao abordar regiões infinitamente distantes de nós, a cosmologia nos obriga a concluir que só saberemos algo sobre estas regiões em nosso futuro, já que atualmente elas estão fora de nosso horizonte. Este conceito, levado ao seu limite, diz que, se uma determinada região do universo enviar um sinal que demore, digamos, um tempo superior ao da existência de nosso Sistema Solar, isto é, se o sinal só chegar a Terra quando o Sol e seus planetas não mais existirem, teremos então que aceitar a ideia de que podem existir regiões do universo suficientemente distantes sobre as quais nós *nunca* obteremos qualquer informação, pelo menos aquelas que viajam à velocidade da luz. Os horizontes implicam, portanto, em regiões que somente tornar-se-ão conhecidas futuramente, ou mesmo em regiões que permanecerão para sempre incognoscíveis. Sendo assim, o que chamamos de universo observável é apenas uma *parte* da totalidade universal. Em outras palavras, somente aquela região do Universo cuja luz *hoje* nos alcança ou nos alcançou no passado.

## 4.3 Singularidades e o Big Bang

O conceito de horizonte discutido acima é intrínseco à cosmologia, indicando portanto que a teoria indica algumas de suas próprias limitações. No entanto, os horizontes não são os únicos conceitos onde a teoria mostra a sua autolimitação. As singularidades, isto é, regiões do espaço-tempo onde as grandezas físicas deixam de ser definíveis, também são limitações da teoria. Exemplos de singularidades são os objetos conhecidos como *buracos negros* e o *big bang*. Ambos aparecem como resultados de aplicações específicas da TRG. Stephen W. Hawking (1942- ) e Roger Penrose (1931- ) estudaram profundamente as singularidades e chegaram a um conjunto de teoremas que afirmam que as singularidades *não* podem ser removidas das teorias gravitacionais geométricas. Portanto, dentro da física que conhecemos não há como evitar que modelos cosmológicos tenham um *big bang* e/ou um *big crunch*. À luz destes teoremas, não nos resta outra opção senão aceitar as singularidades e procurar interpretá-las no contexto das aplicações da TRG.

No caso do *big bang*, ele é uma singularidade que aparece no limite de se olhar ao reverso o processo de expansão do universo. Sua interpretação, porém, é controversa devido às diferentes perspectivas epistemológicas adotadas pelos pesquisadores sobre duas possíveis maneiras de entendermos o que são teorias físicas. Se teorias físicas *correspondem diretamente* à natureza, temos, então, uma identificação entre teoria e objeto. Neste caso se estamos seguros de que há uma expansão e que vários dados relacionados são consistentes com esta ideia, então o big bang *deve* corresponder a algum efeito físico. Resta, portanto, desenvolver a física necessária para finalmente podermos entendê-lo. Se, por outro lado, assumirmos que teorias físicas são *representações* da natureza, então o *big bang* é o local onde a física conhecida encontra os seus limites de validade, isto é, ela é intrinsecamente incapaz de discutir o *big bang*. Nada então podemos afirmar, a não ser que a nossa teoria não é conclusiva. Em outras palavras, concluiríamos que nada podemos concluir. Veremos a seguir que não é possível discutir a natureza do *big bang* sem assumir uma ou outra posição epistemológica. A proposta a ser apresentada a seguir é de que a melhor maneira de entendermos teorias físicas é supondo-as serem representações.

## 5. Teorias Científicas e Representações

Um aspecto fundamental discutido no início desse texto é o de que a ciência só se sente em condições de fazer afirmações em certos domínios de validade. Dito de outra forma, as leis e as teorias científicas referem-se, quando vistas individualmente, a setores específicos do real, sendo, portanto, *representações dos fenômenos naturais*. Esta afirmação reflete a opção epistemológica adotada nesse trabalho, a qual considero a mais conveniente, talvez a única possível, que permite um diálogo tranquilo com outros saberes, particularmente o religioso, sem que ocorra um diálogo de surdos. A tese de que as teorias científicas são representações é antiga, no entanto ficarei aqui circunscrito a apenas um autor da ciência moderna cujas teses são representativas dessa opção epistemológica e fornecem fundamentação suficiente para a discussão da próxima seção onde abordarei o tema acerca do diálogo com outras formas de pensamento.

Ludwig Eduard Boltzmann (1844-1906) foi um notável físico austríaco, um dos grandes físicos do passado, famoso pelas suas contribuições fundamentais à mecânica e termodinâmica estatísticas. Foi o maior defensor da teoria atômica em uma época na qual essa teoria era ainda muito questionada e bastante controversa. Em 1896 Boltzmann escreveu um artigo no qual reafirmava a sua crença de que o atomismo era inevitável nas ciências naturais. Uma das principais razões que motivaram Boltzmann a defender o atomismo era o seu temor de que o acolhimento do energeticismo, teoria rival ao atomismo, por grande parte dos cientistas acarretasse necessariamente a marginalização do atomismo. Nesse caso Boltzmann acreditava que um clima dogmático iria inevitavelmente instalar-se na ciência, o que implicaria na extinção de toda e qualquer possibilidade de progresso científico. Isto porque, tal como no mundo natural, no mundo das teorias científicas, caberia à competição entre teorias desempenhar o papel fomentador do progresso (evolução). É a tentativa de mostrar que uma representação é melhor, ou mais adequada, do que outra que faz com que o cientista a aperfeiçoe.

A principal tese epistemológica adotada por Boltzmann afirmava que *toda teoria científica é uma representação da natureza*. Para Boltzmann aquilo que constitui real e verdadeiramente a natureza é, e permanece para sempre, incognoscível para os cientistas. Além disso, Boltzmann defendia a ideia de que as representações são livres criações dos cientistas, o que o colocava em oposição a alguns de seus contemporâneos que acreditavam que era possível formular descrições diretas daquilo que é percebido com o uso dos nossos sentidos. Ao afirmar que as teorias científicas são criações livres dos cientistas, Boltzmann enfatizava que é impossível o trabalho científico sem o recurso a conceitos teóricos, os quais devem a sua origem ao fato de que é impossível a elaboração de toda e qualquer teoria científica a partir da mera observação dos fatos naturais. Por exemplo, o conceito de atração gravitacional na física de Isaac Newton (1642-1727) resulta da liberdade que os cientistas possuem para representar a natureza, pois pela mera observação da queda dos corpos não é possível elaborar um conceito como este.

Se uma teoria científica representa, ou descreve, a Natureza, não se pode confundir a representação com o objeto representado, visto que são conceitos distintos. Se os confundirmos estaríamos fazendo o mesmo que, por exemplo, identificarmos o desenho de uma laranja com a fruta, o que, obviamente é incorreto. O que uma teoria científica procurar capturar são os aspectos mais básicos, ou importantes, de um fenômeno, da mesma forma que um desenho de uma laranja procura representar esta fruta da melhor forma possível por meio da cor, forma e, conforme a qualidade do desenho, textura, etc. Portanto, da mesma forma que o desenho da laranja tem uma relação com a própria laranja, e representa algumas características reais da fruta, as teorias científicas representam fenômenos e processos que realmente ocorrem na Natureza. Assim, as teorias científicas representam o real, e são, portanto, realistas, ou seja, as teorias científicas não podem conhecer as essências dos fenômenos naturais, suas causas últimas. Essa proposição afirma então que conhecer as essências dos fenômenos naturais é impossível. Além disso, da mesma forma que um desenhista escolhe como, e com que estilo, desenhará as características que ele vê como mais marcantes em uma laranja, essa tese admite implicitamente que os aspectos considerados básicos de uma teoria científica são definidos por *escolha*, individual ou coletiva, e podem mudar com o tempo. Portanto, tais escolhas são feitas em uma determinada época e em um determinado contexto social.

Se nossas teorias científicas são descrições da Natureza, uma consequência importante dessa tese é a de que é necessário aceitar que nossas teorias e modelos possuem limitações intrínsecas e inevitáveis. Por mais elaboradas e detalhadas que sejam, nossas teorias ou modelos jamais serão capazes de representar todas as características da Natureza, da mesma forma que o desenho de uma laranja jamais poderá capturar todos os aspectos da fruta. O desenho de uma laranja é feito em uma folha plana de papel, portanto não pode ser "virado" para podermos olhar do outro lado do desenho da fruta, e é feito a partir de um ponto de vista, de uma perspectiva, da fruta, a qual pode mudar se ao virarmos a laranja descobrirmos uma parte apodrecida no lado oposto a qual não estava representada no desenho original. O outro lado da fruta pode ser representado por *outro* desenho. Em suma, uma teoria científica, assim como um desenho, não pode representar todas as características da Natureza e é, portanto, limitada e incompleta. Existirão então sempre novos aspectos a serem incluídos nas teorias, os quais podem não somente complementar os aspectos já representados, refinando-os, como também podem inclusive modificar radicalmente a imagem que temos da Natureza.

A tese de que as teorias científicas são representações da Natureza permite extrair ainda outra consequência importante: um mesmo fenômeno natural pode ser representado por mais de uma teoria ou modelo. Essa tese é comumente denominada de *pluralismo teórico*. Boltzmann acreditava que os mesmos fenômenos naturais podem ser descritos e explicados de maneiras diferentes a partir da adoção de perspectivas não só distintas e complementares, mas até mesmo excludentes. Assim, a ideia de que as teorias científicas são representações possui também uma implicação pouco comum, que é sua capacidade de *combater o dogmatismo* na medida em que, como dito acima, favorece uma postura pluralista. Diante de uma situação na qual torna-se impossível escolher uma teoria a partir de critérios estritamente científicos, devem os cientistas apelar para outros, situados além dos domínios da ciência. A escolha de uma teoria dependeria de vários fatores, inclusive de preferência epistemológica. Ao final de sua vida, Boltzmann chegou mesmo a afirmar que a escolha por uma teoria científica poderia ser também determinada por critérios pessoais. Assim, até mesmo critérios como moda podem entrar

Uma outra conclusão importante a ser retirada da tese de que as teorias científicas são representações da Natureza é que não pode haver na ciência nenhuma teoria que seja eternamente verdadeira, pois nenhuma teoria científica pode atingir os níveis dos porquês e dos constituintes últimos dos fenômenos naturais. Assim, segue-se que nenhuma teoria científica pode afirmar conhecer verdades imutáveis. *A verdade científica é provisória*. Sendo assim, ela pode ser "alcançada" de diversas maneiras, ou seja, através de teorias diferentes. Quanto maior o número de teorias à disposição dos cientistas, maiores são as chances de se obterem melhores representações dos fenômenos naturais. A tarefa do processo científico é ser uma eterna busca por melhores, mais adequadas, representações da Natureza, sem jamais ser capaz de esgotá-las.

O conhecimento científico é melhor caracterizado por uma busca incessante e sem fim por melhores, mas nunca definitivas, representações dos fenômenos naturais. A substituição de uma teoria científica por outra, característica principal da ciência moderna, obra permanentemente em aberto, só pode acontecer caso assegure-se que nenhuma teoria científica pode alcançar o estágio de definitivamente verdadeira. Em outras palavras, uma teoria científica pode ser melhor do que outra e nada mais do que isso.

# 6. Diálogo com a Religião e outros Saberes

Como afirmei acima, a ciência moderna em geral, e a cosmologia científica em particular, não deveriam se sentir inseguras em dialogar com outras formas de pensamento pois já conquistaram sua autonomia e são suficientemente maduras a ponto de, sob seus próprios critérios de qualidade, abordar temas que interessam a outros saberes. Então, por que deveriam se negar ao diálogo, apenas afirmando de forma arrogante o inegável papel essencial das ciências no mundo moderno? Creio que, mesmo tendo de fato existido no passado, não há hoje mais razão objetiva para tal hostilidade.

No entanto, disposição tranquila para o diálogo não implica que ele existirá ou que, se existir, possa ser fecundo. Para que haja o diálogo é necessário haver um terreno comum, um local que ambos os lados "habitem" e com algo que possa ser compartilhado. A proposição a ser feita aqui é que cabe à filosofia ser o *elemento de ligação* entre, particularmente, mas não exclusivamente, a ciência e a teologia. Com isso quero dizer que a filosofia é a responsável pela existência do diálogo. É da própria natureza da filosofia o estabelecimento deste tipo de diálogo. Basta lembrar o exemplo de Platão que empregou a forma do diálogo para expor as suas ideias e convicções. Foi por meio do diálogo que Sócrates, Platão e Galileu Galilei (1564-1642) seduziram os seus oponentes, tornando-os adeptos de suas teorias.

Precisamos, no entanto, de um último ingrediente para existir esse diálogo, ingrediente este que cabe especificamente à filosofia: mostrar inequivocamente que as teorias científicas são representações do real. No escopo delimitado pela ciência moderna, não há espaço para uma concepção de realismo essencialista como o aristotélico. As essências, aquilo que faz com que o mundo seja o que ele é, não podem ser conhecidas por meio dos procedimentos característicos da ciência moderna. No entanto, nem por isso a ciência moderna se afirma como incapaz de conhecer o real. A ciência moderna nunca abrirá mão de sua pretensão em conhecer o real, pois este é passível de ser parcialmente conhecido empiricamente, o que faz com que possamos sumarizar esse conhecimento por meio da teorização. Em outras palavras, a ciência moderna é realista, pois para ela o real, isto é, a natureza ou mundo externo, existe independentemente do sujeito cognoscente. Assim, o mundo externo existe e existirá independente da percepção e inteligibilidade de qualquer um de nós. O mundo externo não é uma fantasia de nossa mente. Essas duas teses constituem o núcleo central da visão realista. Mas, apesar de ser impossível conhecer o mundo externo inteira e completamente, e certamente ser impossível prever seu futuro corretamente visto que o futuro não está determinado, é útil aprender o que podemos sobre o mundo externo com o objetivo de interpretar melhor a realidade, definida aqui como a interface entre o ser cognoscente e o mundo externo, para melhorar as condições de nossa existência.

Sendo assim, o diálogo só pode ocorrer no campo das representações, conforme as teses epistemológicas de Boltzmann discutidas na seção anterior. Esta é a "arena" do diálogo, onde ideias podem cruzar de um lado para outro e produzir frutos, isto é, novas ideias e aplicações para os lados que dialogam. Mas, cada conceito, cada representação, que se mostrou útil em um determinado domínio de conhecimento somente poderá tornar-se útil em outro se for devidamente adaptado de forma a cumprir os requerimentos e critérios de qualidade próprios de cada domínio. Transposições automáticas jamais darão certo pois cada domínio de conhecimento tem seus próprios critérios e metodologias.

Para exemplificar esse ponto, relembro a passagem acima onde foi dito que o conceito de evolução já era utilizado no século XIX em outras áreas da ciência, mas foi somente no século XX que ele surge na cosmologia científica, tornando-se útil nesta quando foi verificado que ela poderia ser associada aos conceitos produzidos pela física moderna e a TRG para concluir, por exemplo, que existe uma história térmica do universo. Esse simples exemplo mostra que os conceitos, as representações, devem ser devidamente compreendidos dentro de um domínio e, uma vez aceita sua possível utilidade em outro domínio, ser então devidamente interpretadas dentro do último. Abre-se então a possibilidade de que conceitos originados e desenvolvidos em um certo domínio de validade possam ser devidamente reinterpretados e utilizados em outro domínio de validade. Em outras palavras, não há qualquer impedimento que a física "empreste" seus conceitos, suas representações, a outras áreas, como a teologia, ou tome "emprestado" representações, imagens, de outros domínios de conhecimento e os adaptem, os reinterpretem, de forma a se tornarem úteis dentro da física.

Um outro exemplo desse tipo de fertilização cruzada, onde representações originadas em um domínio são reinterpretadas e aplicadas em outro domínio de validade, pode ser visto nas ciências sociais. Em seu livro *"The Uncertainties of Knowledge"*, o sociólogo Immanuel Wallerstein (1930- ) discute o impacto da física Newtoniana na institucionalização das ciências sociais. Em suas palavras,

*"A ciência social foi institucionalizada no final do século XIX sob a sombra da dominação cultural da ciência Newtoniana. (...) Todas as ciências sociais, mas especialmente o trio economia, ciência política e sociologia tornaram-se crescentemente quantitativas e insistiam muito fortemente nas pressuposições de um universo social determinístico. O objeto da ciência social (...) era discernir leis universais similares àquelas que elas acreditavam a física foi capaz de estabelecer"* (Wallerstein 2004, p. 38).

Mais adiante, Wallerstein torna bastante claro como a metodologia teórica da física Newtoniana, ou seja, suas representações, foram vistas como úteis e acabaram sendo reinterpretadas pelas ciências sociais, as quais as adaptaram ao seu próprio contexto:

*"Aqueles que consideravam a ciência social (...) como em busca de leis universais, em geral argumentavam que não havia nenhuma diferença metodológica intrínseca entre o estudo científico dos fenômenos humanos e o estudo científico dos fenômenos físicos."* (Wallerstein 2004, p. 73).

*"... historiadores foram pegos na idolatria da ciência Newtoniana. (...) Eles concebiam os fenômenos sociais como atômicos por natureza. Seus átomos eram os 'fatos' históricos. Esses fatos haviam sido registrados em documentos escritos, em grande parte localizados em arquivos. Eram empiricistas a um grau excessivo."* (Wallerstein 2004, p. 74).

É interessante observar que a crítica de Wallerstein à abordagem metodológica tradicional das ciências sociais, conforme as citações acima, também se baseia em representações originadas na própria física, mas posteriores à física Newtoniana. Em suas palavras,

*"Todos sabemos muito bem que nos últimos 100 anos, e particularmente nos últimos 30 anos, esse modelo Newtoniano da ciência sofreu severo e sustentado ataque vindo de dentro do ventre do próprio monstro, de dentro da própria física e matemática. (...) Eu apenas indicarei os contra-slogans desse ataque: no lugar de certeza, probabilidades; no lugar de determinismo, caos determinístico; no lugar de linearidade, a tendência para se mover longe do equilíbrio em direção à bifurcação; no lugar de dimensões inteiras, fractais; no lugar de reversibilidade, a seta do tempo. E, eu acrescento, no lugar da ciência como fundamentalmente diferente do pensamento humanístico, ciência como parte da cultura".* (Wallerstein 2004, p. 38).

Wallerstein então usa as ideias desenvolvidas pelo físico-químico Ilya Prigogine (1917-2003), a cuja memória ele dedica seu livro, para contrapor a visão sociológica tradicional com as suas teorias baseadas no conceito de sistemas humanos mundiais históricos de longa duração, que evoluem e, ao decaírem, longe do equilíbrio, se bifurcam.

*"Ao levantar a bandeira da 'seta do tempo', ao afirmar que mesmo as menores unidades da matéria física têm uma trajetória histórica, uma que não pode ser ignorada, Prigogine não somente reforçou os cientistas sociais que sempre insistiram que não pode haver análise social que não seja histórica, como também transportou a ciência física para o terreno epistemológico central da ciência social. Ele renovou a clamor por uma ciência unificada".* (Wallerstein 2004, p. 53)

Não se trata aqui de realizar uma exposição das teorias de Wallerstein, cuja leitura é instigante. As citações acima servem apenas como um exemplo a mais de como representações originadas em um domínio de conhecimento podem ser consideradas úteis em outros domínios, gerando fertilização cruzada ao serem devidamente reinterpretadas, adaptadas e conceitualmente incorporadas como representações em áreas por vezes muito distantes de onde elas se originaram.

Para concluir esta seção, é importante observar que debates sobre representações originadas em um domínio e possivelmente utilizadas em outro também ocorrem de forma indireta. Por exemplo, no final do século XX alguns cientistas, principalmente os físicos, começaram a falar em Deus. Alguns deles acreditam que o desenvolvimento da ciência, em particular da cosmologia científica, tornarão possível que os cientistas conheçam a mente de Deus. Embora essas afirmações sejam feitas de modo rápido sem que sejam devidamente apresentadas as razões para acreditarmos que a ciência, e não mais a religião, poderá conhecer os planos que Deus usou para criar a natureza, esse não deixa de ser um fenômeno curioso e até mesmo inesperado. Mas, paralelamente, começaram a aparecer algumas obras escritas por filósofos, teólogos e mesmo por cientistas que analisam de forma mais cautelosa as relações entre a ciência e a religião. Nesses trabalhos não existe a preocupação de verificar se a ciência pode, ou não, descobrir os planos traçados por Deus. Não existe igualmente a preocupação de se criticar a religião em nome da ciência ou vice-versa. Essas análises procuram tomar como ponto de partida resultados científicos e filosóficos onde sobressai-se a tese de que as teorias científicas são representações, além de deixar claro, mesmo de forma indireta, que é preciso deixar para trás os tempos de beligerância entre ciência, filosofia e religião.

Não pretendo realizar aqui uma análise exaustiva da literatura sobre esse assunto, portanto a discussão a seguir será breve e limitada a apenas um autor que considero representativo dessa tendência, o físico, cosmólogo e teólogo jesuíta William R. Stoeger (1943- ). Suas preocupações e posições são semelhantes à nossa discussão acima no que se refere à distinção dos domínios de conhecimento da ciência e da teologia. As posições e reflexões de Stoeger são equilibradas e fecundas, pois ele respeita as particularidades de cada uma das partes envolvidas, e sem esse respeito nada poderá ser alcançado em relação a uma justa apreciação do que cabe à ciência, à filosofia e à religião.

Em suas análises e discussões, Stoeger têm por objetivo encontrar espaço para cada uma das partes envolvidas pois, segundo ele, não existe razão *a priori* para acreditarmos que a ciência ocupou todos os espaços disponíveis para a vida humana. A ciência não deve ser compreendida como um substituto para a religião ou para a filosofia. Por outro lado, a filosofia e a religião devem prestar atenção àquilo que ocorre na ciência; caso contrário, suas análises poderão não ter valor.

Stoeger parte da constatação que, sendo uma teoria científica uma representação da natureza, a ciência não pode afirmar como esta é de fato (ou realmente). Essa tese possibilita a compreensão de que a ciência não conseguirá descrever, os fenômenos vinculados à espiritualidade, pois esta "atua" em um domínio diferente da ciência.

*"O interesse da espiritualidade é muito diferente [daquele da ciência] – é conhecer e responder pessoal e comunalmente a Deus da maneira mais plena possível"*. (Stoeger 2002, p. 90).

A preocupação central de Stoeger vem do fato de que os domínios do espiritual (teologia) e o natural (ciência) são diferentes. Estabelecer essa distinção é a sua primeira preocupação,

*"... vou examinar as leis da natureza – em especial sua posição ontológica – que acredito ser uma questão fundamental que precisa ser solucionada para haver progresso autêntico rumo à integração e para uma rearticulação fecunda da ação divina contra o plano de fundo de nossa descrição científica da natureza e seus processos."* (Stoeger 2002, p. 21).

Portanto, os objetos das ciências naturais não podem ser confundidos com os da religião. Em seguida Stoeger esclarece qual é o domínio de atuação da espiritualidade.

*"Estritamente falando, o conhecimento que se origina da espiritualidade não é tão especializado quanto o que se origina das ciências e é muito mais diretamente pessoal e social em sua origem, em sua relevância e em seu entrosamento (conseqüências). A espiritualidade diz respeito à nossa experiência e nossa resposta a esses elementos que funcionam como absolutos em nossas vidas conscientes, ou revelam o absoluto e dão sentido e orientação fundamentais à maneira como vivemos, como indivíduos e como comunidades"*. (Stoeger 2002, p. 88).

Stoeger então adverte que a teologia e a religião não podem dar as costas aos desenvolvimentos e resultados obtidos pela ciência.

*"... a maneira como interpretamos a Escritura, a filosofia e a teologia hoje indiretamente depende bastante dos avanços em muitas outras disciplinas, até mesmo nas ciências naturais e humanas."* (Stoeger 2002, p. 86)

A afirmação seguinte é bastante esclarecedora da proposta de Stoeger de que a base para o diálogo entre ciência e teologia está no fato de que em qualquer época todas as áreas do conhecimento humano se influenciam, pois participam de uma visão comum de mundo.

*"Os limites de uma área específica – e o enfoque e as bases evidentes a ela apropriados – só são descobertos por meio da interação com outras áreas. Todas as áreas de conhecimento compartilham pelo menos um campo cultural comum e se influenciam mutuamente de diversas formas".* (Stoeger 2002, p. 86).

Stoeger compartilha também da visão de que somente é possível conhecer o que quer que seja por meio da razão humana, sendo que esta está associada a valores, conhecimentos prévios, experiências pessoais e privadas. Esse conjunto de fatores integra o arsenal de instrumentos à disposição do homem para elaborar e validar o conhecimento.

*"É ilusão acreditar que essas representações incrivelmente ricas dos fenômenos são isomorfismos não construídos que meramente descobrimos no mundo real. Ao contrário, são construídos – meticulosamente – e não há indícios de que sejam isomórficos com estruturas no mundo como este é em si".* (Stoeger 2002, p. 36)

Sobre o papel a ser cumprido pela filosofia na intermediação entre a ciência e a teologia, também aos olhos de Stoeger a filosofia é responsável por este diálogo. Espera-se que o conflito resultante seja criativo, o que implica no desdobramento de novos modos de interação e cooperação entre a ciência e a teologia. Além disso, cabe à filosofia a tarefa de avaliar quão "eficiente" é este diálogo e respeitar a complementaridade entre a ciência e a religião, mais especificamente a cosmologia e a teologia. Elas são complementares porque ambas nos fornecem diferentes perspectivas do mesmo tipo de questionamento proveniente das mesmas perguntas advindas da mesma realidade experimentada.

*"Nesse diálogo – e de fato em qualquer reflexão sobre o que ou a cosmologia ou a teologia está fazendo ou é capaz de fazer – a filosofia será um intermediário crucial, tanto ao nível da análise da linguagem comum como ao nível da epistemologia e metafísica. Isto decorre de que aqueles que estão envolvidos na pesquisa em cosmologia e aqueles envolvidos em teologia assumem importantes, embora normalmente inarticuladas, premissas acerca da realidade e nosso conhecimento dela, e ambos usam uma linguagem que reflete essas premissas mais profundas que são assumidas. Todos nós temos uma filosofia implícita. Para que o diálogo entre as duas comunidades seja frutífero, ele precisa abordar bem frequentemente a compatibilidade e aceitabilidade dessas premissas mais básicas do ponto de vista de ambas as disciplinas. Elas precisam ser submetidas a uma crítica permanente."* (Stoeger 1989, p. 2,).

# 7. Conclusão

Nesse trabalho procurou-se apresentar o que é cosmologia, definida como a tentativa de responder a um conjunto de perguntas as quais são comuns a várias formas de pensamento, como a teologia, filosofia e ciência. Mostrou-se que a cosmologia científica moderna é uma forma particular de cosmologia, onde as respostas às perguntas cosmológicas aparecem no contexto da física moderna, em particular da Teoria da Relatividade Geral de Einstein. Mostrou-se também que a cosmologia científica tem seus próprios pressupostos e arcabouço teórico, que parte da premissa de que o Universo pode ser estudado como um sistema físico único e define a totalidade como sendo o conjunto de todos os objetos que nos cercam e que se influenciam mutuamente. Foram apresentadas as evidências observacionais que fornecem suporte empírico às conclusões teóricas da cosmologia científica e como esta interpreta conceitos como origem, universo observável, *big bang*, etc.

A seguir apresentei a visão epistemológica adotada nesse trabalho que, resumidamente, assume que as teorias científicas não passam de representações da natureza, do real, ou mundo externo e, portanto, são intrinsecamente limitas e incompletas, o que abre a possibilidade de haver uma multiplicidade de representações, configurando o que comumente se chama de pluralismo teórico. Foi argumentado então que é no terreno das representações que pode existir um diálogo entre as várias formas de pensamento, não só o científico e religioso, mas também entre os vários domínios de pensamento científico, e que conceitos originados em um domínio podem ser apropriados e utilizados em outros, mas que isso somente funciona se não houver transposição automática, mas tendo clareza do significado dos conceitos dentro de seus domínios de origem e respeitando os critérios de qualidade e metodologia de cada domínio de pensamento.

Para concluir, acredito que não é demais enfatizar que as questões relativas à cosmologia científica não lhe pertencem somente, pois, afinal, aquele que em tempos muito remotos, perguntou-se pela primeira vez *"De onde viemos?"*, *"Qual a origem desse todo que me cerca?"* e *"Qual o destino desse Universo?"* com certeza não fez essas perguntas motivado por questões relacionadas à física do Universo. Com certeza a sua preocupação era conhecer os desejos e desígnios de seus Deuses.

# 7. Bibliografia

.